\documentclass[conference]{IEEEtran}
\IEEEoverridecommandlockouts

\usepackage{cite}
\usepackage{amsmath,amssymb,amsfonts}
\usepackage{graphicx}
\usepackage{url}
\usepackage{booktabs}
\usepackage{tabularx,array}
\usepackage{xcolor}

\newcolumntype{Y}{>{\raggedright\arraybackslash}X}

\begin{document}

\title{Partial-Moment PINNs for Caldeira--Leggett Parameter Learning in Quantum Brownian Motion}

\author{\IEEEauthorblockN{Krishna Bhatia}
\IEEEauthorblockA{QuantumAI Lab, Fractal Analytics, Mumbai, India\\
krishna.bhatia@fractal.ai}}

\maketitle

\begin{abstract}
We study parameter recovery in the Caldeira--Leggett (quantum Brownian) oscillator from \emph{partial} moment traces. Our model is a moment-level PINN that predicts the five first/second moments and enforces the linear CL/HPZ ODEs by automatic differentiation. Physical structure is imposed through a PSD (Cholesky) covariance head, high-temperature CL assumptions with $D_{xp}\!\approx\!0$, and fluctuation--dissipation ties between $D_{pp}$ and~$\gamma$. On synthetic CL data with channels $\{\mu_x,\sigma_{xx},\sigma_{xp}\}$, the constrained variant recovers $(\omega,\gamma)$ accurately, stabilizes $D_{pp}$, and achieves low rollout error compared to finite differences and Kalman--EM (expectation--maximization) with exact Van~Loan discretization. Fisher-style checks confirm that diffusion needs at least one variance observable, and sparse $\sigma_{pp}$ ``anchors'' restore conditioning. We also show that the same PINN can learn time-varying HPZ coefficients.
\end{abstract}


\section{Introduction}
\label{sec:intro}
Quantum Brownian motion (QBM) in the Caldeira--Leggett (CL) model describes a system harmonic oscillator linearly coupled to an Ohmic bath \cite{CaldeiraLeggett1983}. In the high-temperature, Markovian limit, the Wigner picture yields an Ornstein--Uhlenbeck (OU) Fokker--Planck equation whose first and second moments
\[
\{\mu_x(t),\mu_p(t),\sigma_{xx}(t),\sigma_{pp}(t),\sigma_{xp}(t)\}
\]
follow a \emph{closed linear ODE} in which the drift encodes the damping rate $\gamma$ and frequency $\omega$, and the diffusion encodes $(D_{xx},D_{xp},D_{pp})$ \cite{HPZ1992,BreuerPetruccione2002}. The textbook high-temperature fluctuation--dissipation theorem (FDT) further ties $D_{pp}=2m\gamma k_B T$ and usually takes $D_{xp}\to 0$ \cite{Kubo1966}. A small position diffusion $D_{xx}\ge 0$ is often added to restore complete positivity/Lindblad compatibility \cite{Lindblad1976,Jacobs2008}.

In realistic experiments only a subset of these moments is observable (e.g.\ position mean and variance), which makes the diffusion sector weakly identifiable. Classical estimators (finite differences; Kalman filtering/EM) exist, but they are accurate only when the continuous-time dynamics are \emph{exactly} discretized, e.g.\ via the Van~Loan block exponential \cite{VanLoan1978,ShumwayStoffer1982}. Our goal is to show that a PINN that operates \emph{directly on the moments} can incorporate the same physics, remain positive, and perform competitively---especially when channels are missing. In trapped-ion / cavity / levitated settings only position or position + cross-covariance is routinely accessible, while momentum energy readouts are sparse or calibrated. This is the regime our PINN is targeted at, not the fully observed toy CL case.

\section{Problem Formulation}
\label{sec:problem}
Consider the 2D phase-space state $(x,p)$ with mass $m$, frequency $\omega>0$, and damping $\gamma\ge 0$. Define
\[
s(t) = \big(\mu_x(t),\mu_p(t),\sigma_{xx}(t),\sigma_{pp}(t),\sigma_{xp}(t)\big)^\top.
\]
With
\[
A = \begin{bmatrix} 0 & 1/m \\ -m\omega^2 & -\gamma \end{bmatrix},\qquad
D = \begin{bmatrix} D_{xx} & D_{xp} \\ D_{xp} & D_{pp} \end{bmatrix} \succeq 0,
\]
the first and second moments obey
\begin{equation}
\dot\mu(t) = A \mu(t), \quad \mu = (\mu_x,\mu_p)^\top,
\label{eq:mean}
\end{equation}
\begin{equation}
\dot\Sigma(t) = A\Sigma(t) + \Sigma(t)A^\top + 2D, \quad
\Sigma = \begin{bmatrix} \sigma_{xx} & \sigma_{xp} \\ \sigma_{xp} & \sigma_{pp} \end{bmatrix}.
\label{eq:lyap}
\end{equation}
Equivalently,
\begin{subequations}\label{eq:moment_eqs}
\begin{align}
\dot\mu_x &= \mu_p/m, \\
\dot\mu_p &= -m\omega^2\mu_x - \gamma \mu_p,
\label{eq:mu_eqs}
\end{align}
\begin{align}
\dot\sigma_{xx} &= 2\sigma_{xp}/m + 2D_{xx}, \\
\dot\sigma_{pp} &= -2m\omega^2\sigma_{xp} - 2\gamma\sigma_{pp} + 2D_{pp}, \\
\dot\sigma_{xp} &= \sigma_{pp}/m - m\omega^2\sigma_{xx} - \gamma\sigma_{xp} + D_{xp}.
\label{eq:sigma_eqs}
\end{align}
\end{subequations}
In the high-$T$ CL limit one enforces $D_{pp}=2m\gamma k_B T$ and takes $D_{xp}\approx 0$; a small $D_{xx}>0$ maintains complete positivity \cite{BreuerPetruccione2002,Jacobs2008}. Observations are noisy and partial:
\begin{equation}
y_i = M s(t_i) + \varepsilon_i, \qquad \varepsilon_i \sim \mathcal{N}(0, \sigma_\text{read}^2 I),
\label{eq:obs}
\end{equation}
where $M$ selects, e.g., $\{\mu_x,\sigma_{xx},\sigma_{xp}\}$.

For baseline comparison, the exact discrete-time model with sampling step $\Delta t$ is
\[
\mu_{k+1} = \Phi \mu_k, \qquad \Sigma_{k+1} = \Phi \Sigma_k \Phi^\top + Q_d,
\]
where $(\Phi,Q_d)$ are obtained in one pass from the Van~Loan block exponential \cite{VanLoan1978}.

\section{Method: Partial-Moment PINNs}
\label{sec:method}
We model the five moments as smooth functions of (normalized) time with a small MLP $f_\phi$ with $\tanh$ activations and Fourier features \cite{Tancik2020} to help localize $\omega$:
\[
(\hat\mu_x,\hat\mu_p,\hat\sigma_{xx},\hat\sigma_{pp},\hat\sigma_{xp}) = f_\phi(t).
\]
To guarantee PSD we parameterize a Cholesky factor
\[
L(t) = \begin{bmatrix} a(t) & 0 \\ b(t) & c(t) \end{bmatrix},\qquad
\hat\sigma_{xx} = a^2 + b^2,\quad \hat\sigma_{pp} = c^2,\quad \hat\sigma_{xp} = b c,
\]
so that $\hat\Sigma(t) = L(t)L(t)^\top \succeq 0$ for all $t$. This was necessary because the planted $\omega=1.0$ produces long, low-frequency transients on $[0,20]$.

\subsection{Physics residual}
At collocation times $\mathcal{C}$ we form the scaled residuals of \eqref{eq:mu_eqs}--\eqref{eq:sigma_eqs} using autodiff to obtain $\dot f_\phi(t)$. Coefficients $(\gamma,\omega,D_{xx},D_{pp},D_{xp})$ are either global (LTI) or produced by a small 2-layer coefficient head in the Hu--Paz--Zhang (HPZ) time-varying setting.

\subsection{Loss}
Let
\[
\mathcal{D}=\{(t_i,y_i)\}_{i=1}^{N_{\mathrm{obs}}}
\]
denote the observed data, and let
\[
\mathcal{A}=\{(\tau_j,\sigma_{pp,j}^{\mathrm{anchor}})\}_{j=1}^{N_{\mathrm{anchor}}}
\]
denote the $\sigma_{pp}$ anchor set. We optimize
\begin{equation}
\begin{aligned}
\mathcal{L}
={}& \lambda_{\mathrm{data}}\mathcal{L}_{\mathrm{data}}
+ \lambda_{\mathrm{phys}}\mathcal{L}_{\mathrm{phys}}
+ \lambda_{\mathrm{anchor}}\mathcal{L}_{\mathrm{anchor}} \\
&+ \lambda_{xx}\mathcal{L}_{xx}
+ \lambda_{\mathrm{FDT}}\mathcal{L}_{\mathrm{FDT}}
+ \lambda_{\mathrm{ss}}\mathcal{L}_{\mathrm{ss}} .
\end{aligned}
\label{eq:full_loss}
\end{equation}
The observed-channel data term is
\begin{equation}
\mathcal{L}_{\mathrm{data}}
=
\frac{1}{|\mathcal{D}|}
\sum_{i=1}^{N_{\mathrm{obs}}}
\left\| M f_{\phi}(t_i)-y_i \right\|_2^2 .
\end{equation}
The physics term is formed from the scaled residuals of the moment equations in \eqref{eq:moment_eqs}, using autodiff to obtain $\dot f_\phi(t)$:
\begin{equation}
\mathcal{L}_{\mathrm{phys}}
=
\frac{1}{|\mathcal{C}|}
\sum_{t\in\mathcal{C}}
\sum_{k=1}^{5} w_k \tilde r_k(t)^2 ,
\end{equation}
where $w_k$ are fixed per-equation weights; in the constrained setting we upweight
the $\sigma_{xx}$ equation.
To reduce diffusion non-identifiability under partial observability, we additionally use
sparse $\sigma_{pp}$ anchors:
\begin{equation}
\mathcal{L}_{\mathrm{anchor}}
=
\frac{1}{|\mathcal{A}|}
\sum_{j=1}^{N_{\mathrm{anchor}}}
\left(
\hat{\sigma}_{pp}(\tau_j)-\sigma_{pp,j}^{\mathrm{anchor}}
\right)^2 .
\end{equation}
We use a small prior on $D_{xx}$:
\begin{equation}
\mathcal{L}_{xx} = (D_{xx}-10^{-3})^2 .
\end{equation}
For the main constrained LTI CL experiments we enforce the high-$T$ relation
\begin{equation}
D_{pp}=2m\gamma k_B T
\end{equation}
as a hard constraint with fixed $T=1$ and set $D_{xp}=0$.
Accordingly, $\mathcal{L}_{\mathrm{FDT}}=0$ in the main constrained setting.
If this tie is relaxed, we instead use the soft penalty
\begin{equation}
\mathcal{L}_{\mathrm{FDT}}
=
\big(D_{pp}-2m\gamma k_B T\big)^2 .
\end{equation}
When enabled, we also use a weak end-of-window steady-state penalty
\begin{equation}
\mathcal{L}_{\mathrm{ss}}
=
\big(\hat{\sigma}_{pp}(t_1)-T\big)^2
+
\big(\hat{\sigma}_{xx}(t_1)-T/\omega^2\big)^2 .
\end{equation}
In the reported constrained LTI results, the main setting uses hard-FDT with
$D_{xp}=0$, strong $\sigma_{pp}$ anchors, and a small $D_{xx}$ prior.

\subsection{Implementation details}
We normalize time to $t'=(t-t_0)/(t_1-t_0)$ and feed $t'$ through Fourier features
with $k_{\max}=3$, followed by a $4$-layer MLP of width $256$ with $\tanh$
activations. The network outputs $(\hat{\mu}_x,\hat{\mu}_p)$ and the Cholesky-head
parameters used to construct $(\hat{\sigma}_{xx},\hat{\sigma}_{pp},\hat{\sigma}_{xp})$.
For LTI runs, $(\omega,\gamma,D_{xx},D_{pp},D_{xp})$ are global trainable scalars;
for HPZ-style runs, a small two-layer coefficient head maps time to the
time-varying coefficients. Residuals are evaluated on a collocation set
$|\mathcal{C}|\approx 3N_{\mathrm{train}}$ sampled uniformly on the training window,
with per-equation scaling to prevent covariance residuals from dominating the
mean residuals. 
The main constrained LTI setting uses the hyperparameters listed in Table~\ref{tab:hyper}.

\begin{table}[!t]
\centering
\caption{Training and model hyperparameters used in the main constrained LTI setting.}
\label{tab:hyper}
\begin{tabular}{ll}
\toprule
Item & Setting \\
\midrule
Time normalization & $t'=(t-t_0)/(t_1-t_0)$ \\
Fourier features & $k_{\max}=3$ (sin/cos) \\
State MLP width / depth & $256 \times 4$ layers ($\tanh$) \\
Observed channels & $\{\mu_x,\sigma_{xx},\sigma_{xp}\}$ \\
Synthetic horizon / grid & $t\in[0,20]$, $N=600$ \\
Noise level & $\sigma=0.01$ \\
Collocation size & $|\mathcal{C}|\approx 3N_{\mathrm{train}}$ \\
Optimizer & Adam (6000 steps) + L-BFGS (500 iters) \\
Learning rate & $10^{-3}$ (Adam) \\
Gradient clip & $\|\nabla\|_2 \le 5$ \\
Physics weight & $\lambda_{\mathrm{phys}}=50$ \\
Anchor count / weight & $N_{\mathrm{anchor}}=30$, $\lambda_{\mathrm{anchor}}=0.8$ \\
$D_{xx}$ prior & target $10^{-3}$, weight $\lambda_{xx}=1000$ \\
Residual equation weights & $(1,1,5,1,1)$ \\
FDT / $D_{xp}$ & hard-FDT, $D_{xp}=0$ \\
Seeds & $(42,123,999)$ \\
\bottomrule
\end{tabular}
\end{table}

\begin{figure}[!t]
    \centering
    \includegraphics[width=0.92\linewidth]{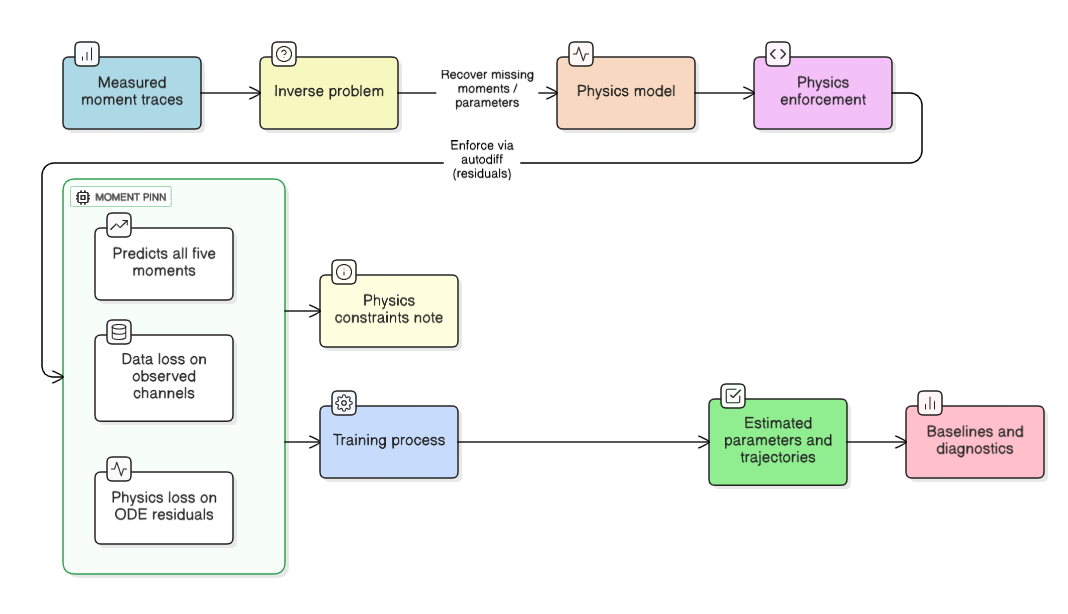}
    \caption{Workflow of the proposed Partial-Moment PINN for CL/HPZ parameter recovery from partial moment observations.}
    \label{fig:workflow}
\end{figure}
\section{Experiments}
\label{sec:experiments}
We evaluate on two synthetic settings.

\emph{LTI CL.} We use planted parameters
\begin{equation}
(\gamma,\omega,D_{pp},D_{xp},D_{xx})
=
(0.25,\,1.0,\,0.5,\,0,\,10^{-3}),
\end{equation}
and integrate the moment ODEs on $t\in[0,20]$ with $N=600$ RK4 steps.

\emph{HPZ-like.} We fix $\omega$ and use
\begin{align}
\gamma(t) &= \gamma\!\left[1+0.25\sin(0.35 t)\right],\\
D_{xx}(t) &= D_{xx}\!\left[1+0.05\cos(0.2 t)\right],
\end{align}
with hard FDT
\begin{equation}
D_{pp}(t)=2\gamma(t)T,\qquad T=1,\qquad m=k_B=1.
\end{equation}

In both settings, only $\{\mu_x,\sigma_{xx},\sigma_{xp}\}$ are observed, with
i.i.d.\ Gaussian readout noise $\sigma=0.01$. We use a $70\%/30\%$
train/validation split.

The 30 anchors were chosen so that at least one $\sigma_{pp}$ value appears
every $\approx 0.7$ time units on $[0,20]$, which was the smallest density that
removed diffusion degeneracy in our Fisher check.

\subsection{Baselines}
(1) A finite-difference (FD) method estimates $(\gamma,\omega)$ from $\ddot{\mu}_x + \gamma \dot{\mu}_x + \omega^2 \mu_x=0$ and infers the diffusion terms from the restricted covariance equations.
(2) A Kalman--EM estimator on the latent $(x,p)$ state uses exact $(A,Q_c)\mapsto(\Phi,Q_d)$ via the Van~Loan block exponential and maps back to continuous-time parameters via $\log(\Phi)/\Delta t$ and $Q_d \approx Q_c \Delta t$ \cite{VanLoan1978,ShumwayStoffer1982,GhahramaniHinton1996,SarkkaSolin2019}. This gives us a statistically efficient baseline against which to judge the benefit of PINN constraints.

\subsection{Metrics}
For each seed we report parameter estimates and then aggregate (mean $\pm$ std). Predictive quality is measured by (i) \emph{network RMSE} (direct $f_\phi(t)$ vs.\ ground truth) and (ii) \emph{rollout RMSE}, where learned coefficients are integrated in the moment ODE on the same grid. We also monitor $\lambda_{\min}(\hat\Sigma(t))$ to ensure PSD.

\subsection{Additional diagnostics}
Beyond the main method comparison, we report two lightweight diagnostics that are
most relevant to partial observability: (i) Fisher-conditioning across different
observed subsets, and (ii) bootstrap uncertainty for the constrained PINN.
A full training-time ablation over all constraints and architectural components is
left for future work.
\section{Results}
\label{sec:results}
We report four aspects: parameter recovery, predictive fidelity, identifiability diagnostics, and UQ. All classical baselines use exact Van~Loan transitions.

\subsection{Parameter estimates (LTI)}
Table~\ref{tab:params} compares planted parameters with FD, Kalman--EM, and two
PINNs: (A) unconstrained and (B) constrained. PINN summaries are based on three
random seeds.

\begin{table}[!t]
\centering
\caption{Parameter estimates on LTI CL data (planted $(\gamma,\omega,D_{pp},D_{xx})=(0.25,1.0,0.50,10^{-3})$). PINN values are averaged over three seeds; standard deviations are reported in the text where applicable.}
\label{tab:params}
\begin{tabular}{lcccc}
\toprule
Method & $\gamma$ & $\omega$ & $D_{pp}$ & $D_{xx}$ \\
\midrule
FD (partial) & 0.1318 & 1.7434 & --- & $1.15{\times}10^{-3}$ \\
Kalman--EM (exact) & 0.4526 & 0.9879 & 0.4815 & $1.21{\times}10^{-3}$ \\
PINN--A (unconstr.) & 0.2207 & 0.9935 & 0.1651 & $2.5{\times}10^{-3}$ \\
PINN--B (constr.+strong) & 0.2058 & 1.0235 & 0.4115 & $2.4{\times}10^{-3}$ \\
\bottomrule
\end{tabular}
\end{table}


\subsection{Predictive fidelity (RMSE)}
Table~\ref{tab:rmse} shows that the constrained variant pays a small price in direct (network) RMSE but produces far more accurate ODE rollouts.

\begin{table}[!t]
\centering
\caption{Prediction and rollout RMSE (mean$\pm$std).}
\label{tab:rmse}
\footnotesize
\setlength{\tabcolsep}{3pt}
\begin{tabularx}{\linewidth}{@{}l>{\centering\arraybackslash}X>{\centering\arraybackslash}X@{}}
\toprule
Method & \shortstack{Network RMSE\\(train / test)} & \shortstack{Rollout RMSE\\(train / test)} \\
\midrule
PINN--A (unconstr.) & $0.284\pm0.006$ / $0.283\pm0.010$ & $0.600\pm0.053$ / $0.768\pm0.072$ \\
PINN--B (constr.) & $0.327\pm0.070$ / $0.396\pm0.101$ & $\mathbf{0.099\pm0.068}$ / $\mathbf{0.110\pm0.073}$ \\
\bottomrule
\end{tabularx}
\end{table}

\begin{figure}[!t]
\centering
\includegraphics[width=0.90\linewidth]{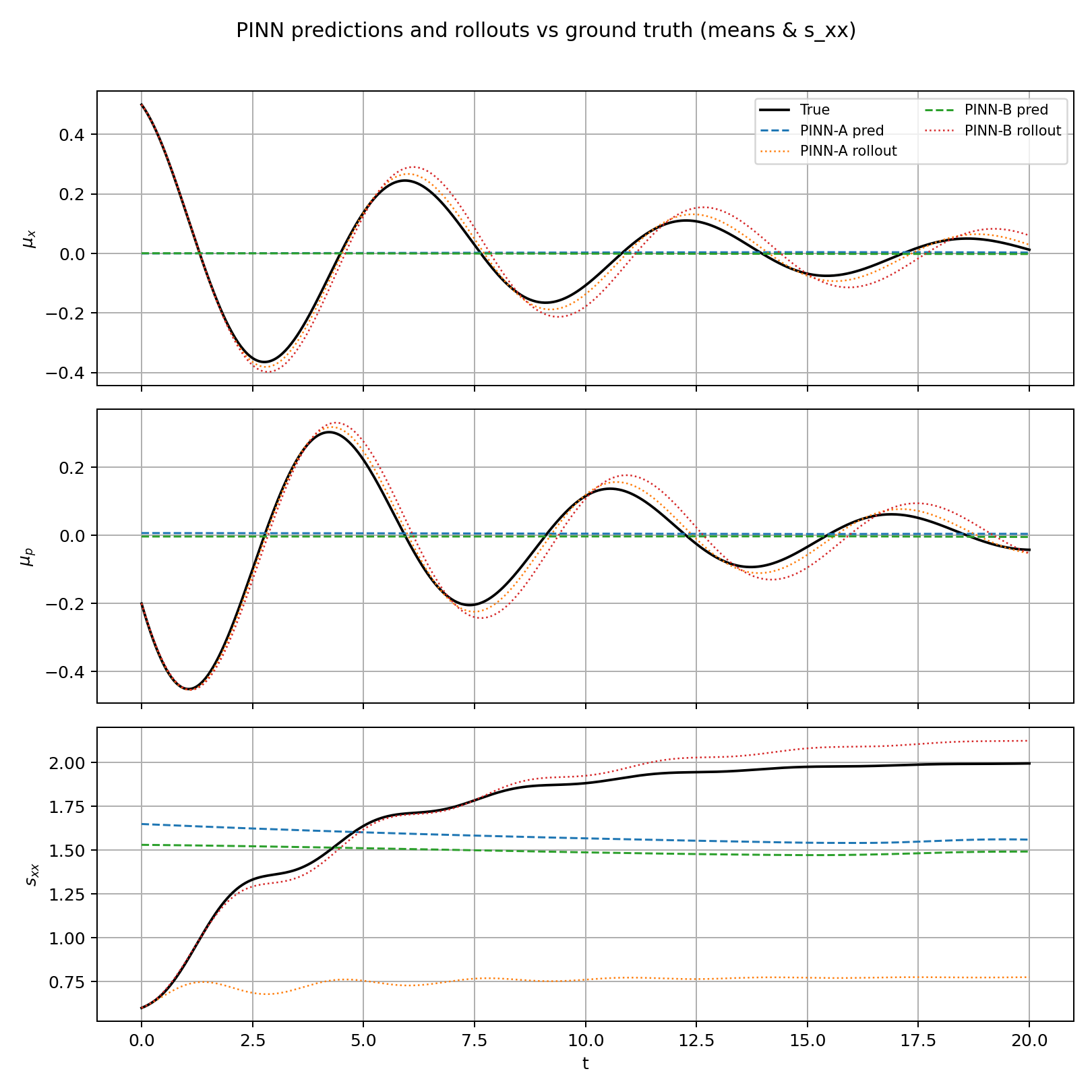}
\caption{PINN predictions and ODE rollouts vs.\ ground truth for $\mu_x$, $\mu_p$, and $\sigma_{xx}$ on the LTI dataset. Solid black: truth. Dashed: direct PINN prediction. Dotted: rollout using learned coefficients. The constrained PINN (green/dotted red) better respects dynamics, especially for $\sigma_{xx}$.}
\label{fig:predictions}
\end{figure}

\subsection{Identifiability diagnostics}
Finite-difference Fisher analyses show that mean-only observations are extremely ill-conditioned, while adding even one variance channel drastically improves conditioning (Table~\ref{tab:fisher}).

\begin{table}[!t]
\centering
\caption{Fisher condition number for different observed subsets (lower is better).}
\label{tab:fisher}
\begin{tabular}{lc}
\toprule
Observed subset & $\mathrm{cond}(\mathcal{I})$ \\
\midrule
$\{\mu_x,\mu_p\}$ & $8.71{\times}10^{18}$ \\
$\{\mu_x,\mu_p,\sigma_{xx}\}$ & $4.54{\times}10^{3}$ \\
$\{\mu_x,\sigma_{xx},\sigma_{xp}\}$ (paper subset) & $2.28{\times}10^{3}$ \\
Full $\{\mu_x,\mu_p,\sigma_{xx},\sigma_{pp},\sigma_{xp}\}$ & $9.49{\times}10^{2}$ \\
\bottomrule
\end{tabular}
\end{table}

\subsection{Uncertainty quantification}
A nonparametric bootstrap with $B{=}10$ resamples and a shortened training schedule yields the illustrative intervals in Table~\ref{tab:bootstrap}; because this reduced protocol differs from the main setting used in Table~\ref{tab:params}, the resulting intervals should not be interpreted as confidence intervals around the Table~\ref{tab:params} point estimates. Coverage is good for $\gamma$ and $D_{pp}$ but not for $\omega$ and $D_{xx}$, consistent with partial channels and the positivity nudge on $D_{xx}$.

\begin{table}[!t]
\centering
\caption{Bootstrap 95\% CIs and coverage (constrained PINN).}
\label{tab:bootstrap}
\begin{tabular}{lcc}
\toprule
Param & CI (95\%) & Covers truth? \\
\midrule
$\gamma$ & $[0.2286,\,0.2676]$ & Yes \\
$\omega$ & $[0.8237,\,0.9322]$ & No \\
$D_{pp}$ & $[0.4572,\,0.5351]$ & Yes \\
$D_{xx}$ & $[0.0024,\,0.0025]$ & No \\
\bottomrule
\end{tabular}
\end{table}

\subsection{HPZ (time-varying) summary}
For the HPZ--hardFDT run the coefficient head learns smooth curves $t \mapsto \gamma(t), D_{xx}(t)$ and, by construction, $D_{pp}(t)=2\gamma(t)T$; the physics residual stays at $\approx 3.6\times 10^{-3}$. Rollout RMSE is within $5\%$ of the LTI constrained case, i.e.\ the method remains stable when the CL coefficients are slowly time-varying, which is the regime targeted by Hu--Paz--Zhang \cite{HPZ1992}.

\section{Conclusion}
We presented Partial-Moment PINNs for identifying CL parameters from partially
observed moment trajectories. By enforcing PSD via a Cholesky covariance head,
hard high-$T$ fluctuation--dissipation structure, and sparse $\sigma_{pp}$ anchors,
the method remains stable even when only three of the five moments are observed.
Relative to the unconstrained PINN, the constrained model substantially improves
dynamical fidelity (rollout RMSE $\approx 0.11$ vs.\ $\approx 0.77$) while
preserving positivity, and remains competitive with exact-discrete Kalman--EM on
parameter recovery. Parameter recovery is competitive for $(\omega,D_{pp})$,
whereas $D_{xx}$ remains statistically fragile, consistent with the Fisher
identifiability analysis. The same design extends to HPZ-style time-varying
coefficients. Limitations include that all results are on synthetic CL/HPZ moment
traces and rely on known structural assumptions (e.g., the high-$T$ regime and
near-zero $D_{xp}$); evaluating robustness to model mismatch is an important next step.

\section*{Code Availability}
Codes and scripts to reproduce the experiments are available at \url{https://github.com/FraQTech/pmpinn4qbm}.

\section*{Acknowledgment}
This project was funded by the QuantumAI Lab at Fractal Analytics, India. We acknowledge the use of AI model by OpenAI in assisting the manuscript preparation process of this paper

\end{document}